\documentclass[namedreferences,hyperref,optionalrh]{spr-sola}
\usepackage{graphicx}        % For eps figures, newer & more powerfull
\usepackage{color}           % For color text: \color command
\usepackage{lmodern}

\chardef\us=`\_

\begin{document}

\begin{frontmatter}
\title{Alfv\'en Waves in Partially Ionised Solar Steady-State Plasmas}

\author[addressref={aff1,aff2},corref,email={nfalshehri1@sheffield.ac.uk}]{\inits{N.F.}\fnm{Nada F.}~\snm{Alshehri}\orcid{0009-0008-1922-9581}}
\author[addressref={aff1},email={i.ballai@sheffield.ac.uk}]{\inits{I.B.}\fnm{I.}~\snm{Ballai}\orcid{123-456-7890}}
\author[addressref={aff1},email=
{g.verth@sheffield.ac.uk}]{\inits{G.V.}\fnm{G.}~\snm{Verth}\orcid{987-654-3210}}
\author[addressref=aff3,email={v.fedun@sheffield.ac.uk}]{\inits{V.F.}\fnm{V.}~\snm{Fedun}\orcid{0000-0002-0893-7346}}

\address[id=aff1]{Plasma Dynamics Group, School of Mathematical and Physical Sciences,\\
The University of Sheffield, Hicks Building, Hounsfield Road,\\
Sheffield, S3 7RH, United Kingdom}

\address[id=aff2]{Department of Mathematics, College of Sciences, King Khalid University, Abha 61413, Saudi Arabia}

\address[id=aff3]{Plasma Dynamics Group, Department of Automatic Control and Systems Engineering,\\
The University of Sheffield, Mappin Street,\\
Sheffield, S1 3JD, United Kingdom}

\runningauthor{Alshehri et al.}
\runningtitle{Alfv\'en Waves in Partially Ionised Steady-State Plasmas}

\begin{abstract}

Our study investigates the properties of Alfv\'en waves in partially ionised solar plasmas in the presence of steady, field-aligned, flows of charged and neutral particles. Our work aims to understand how such flows modify wave propagation and damping in environments where ion-neutral collisions are significant. We employ a two-fluid model that treats ions and neutrals as separate, colliding fluids and incorporates background steady flows for both species. Using a combination of analytical dispersion analysis and numerical solutions, we examine the impact of these flows on the behaviour of Alfv\'en waves. Our results show that steady flows lead to substantial modifications of wave properties, including Doppler shifts, propagation direction reversal, flow-dependent changes in damping rates, and the appearance of a new mode associated with neutral flow and collisional coupling. We also identify conditions under which flow-driven mode conversion can arise. Our results offer new insights into the interplay between plasma flows and particle collisions in the regions of the solar atmosphere where partial ionisation is relevant. 

%\texttt{sola\us keyword\us list.txt}.  
\end{abstract}
\keywords{Alfvén waves- Magnetohydrodynamics (MHD)-Plasma flows- Ion-Neutral Collisions- Wave damping - Stream instabilities- Solar: atmosphere}
\end{frontmatter}
%Online-only material: color figures
%-------------------------------------------------

\section{Introduction}
     \label{S-Introduction} 

The importance of magnetohydrodynamic (MHD) waves in partially ionised plasmas for energy dissipation and plasma dynamics has led to an increased interest in the properties of these waves over recent years. In contrast to fully ionised plasmas (e.g. the solar corona), partially ionised plasmas display a string of distinct physical effects like ion-neutral frictional heating, collisional heat transfer, charge exchange, peculiar transport mechanisms such as Cowling's resistivity, and isotropic thermal conduction by neutrals. These effects are all crucial for a thorough understanding of astrophysical plasmas in a variety of environments \cite{ballester2018partially}. Recent research revealed that in the low solar atmosphere (photosphere and chromosphere), ion-neutral collisions represent a very effective energy dissipation mechanism that can lead to plasma heating \citep[e.g.,][]{khomenko2012heating,martinez2012two,russell2013propagation,leake2014ionized,mcmurdo2023phase}. The framework that can cast the effects due to collisions between particles requires a multi-fluid approach, where each of the constituent fluid is treated separately, and the governing equations of each fluid are coupled to each other through terms that describe the collisional transfer of energy and momentum between particles \citep{khomenko2014fluid,hunana2022generalized}. In practice, in the solar chromosphere, charged particles (electrons and protons in the case of a pure hydrogen plasma) are strongly coupled via long-range electrostatic collisions, and, therefore, it is customary to treat the plasma as a mixture of charged particles and neutrals. In this case, the temporal scales of dynamical processes (waves, in our case) of interest are of the order of the collisional frequency of particles. Although in principle, all charged particles can collide with neutrals, the collisions between electrons and neutrals are usually neglected, as the great mass difference between them leads to an insignificant transfer of energy or momentum. The theory behind waves in a two-fluid plasma is well established \citep[see, e.g.,][]{Zaqarashvili2011,soler2013alfven, Soler2fluid,Alharbi2fluid,Kumar2024}.

%The single-fluid model works well when the wave frequency is significantly lower than the frequency at which charged and neutral particles interact. A more complete framework is offered by the two-fluid model, which captures regimes in which the wave frequency is lesser than the ion-electron collision frequency but equal to or higher than the ion-neutral collision frequency. In situations with even higher frequencies or more complexity, a kinetic or three-fluid model—which treats ions, electrons, and neutrals as separate fluids might be needed \citep{soler2024magnetohydrodynamic}.

The behavior of Alfv\'en waves in partially ionized plasmas is significantly more complex than that of their counterparts in fully ionized plasmas, where Alfv\'en waves can be described by the set of ideal MHD equations, and the magnetic field is frozen into the fluid. In contrast, dynamics in partially ionised plasmas require dissipative equations arising from collisional interaction between species. In partially ionised plasmas, Alfv\'en waves undergo natural damping, making them an ideal candidate to explain plasma heating \citep{Soler2016,Soler2017,Kuzma2020,Melis2023,mcmurdo2023phase,Kumar2024}. Waves are important not only to explain plasma heating but also to serve as tools for diagnosing the magnetic field or the ionisation state of the plasma. In this respect, torsional Alfv\'en waves observed in solar prominence fibrils \citep{Kohutova2020} were used in the study by \cite{Ballai2020} to determine the ionisation degree of the plasma. The effect of background flows on the stability of various configurations in partially ionised plasma has been investigated by several authors  \cite{Soler_2012khi,Ballai2015,Martinez2015,Melis2025}

Our study examines the properties of Alfv\'en waves propagating in a partially ionised plasma in the presence of a differentiated field-aligned steady-state flow for the two species of particles. Our analysis builds on the earlier research by \cite{soler2013alfven} that investigated the propagation of Alfv\'en waves in a partly ionised two-fluid plasma for a static equilibrium. In a partially ionised plasma, when particles are not strongly coupled to each other, different species may flow at different speeds. In the solar photosphere, the plasma is very weakly ionised, and the dynamics is often driven by convection. At the same time, the photosphere is permeated by a weak magnetic field, forming the photospheric carpet. Given its very low ionisation degree, the photospheric plasma is moved by pressure forces, and the motion of neutrals drags the small amount of charged particles with them. As charges are still coupled to the magnetic field, this results in a drift of the neutral and charged species, resulting in different speeds at which they move. In the higher region of the partially ionised solar atmosphere, the collisions are not so frequent, and it is natural to consider that the motion of low-density neutral species is different from the motion of charged particles that keep gyrating around the background magnetic field; therefore, it is again natural to imagine that the two species possess different speeds. Interestingly, several studies have evidenced clear differences in the Doppler velocity of ions and neutrals \citep{Khomenko2016,Anan2017,Stellmacher2017,Zapior2022,Gonzales2024,Hillier2024}.

Our paper is structured as follows: in Section \ref{S-aug} we introduce the mathematical background and derive the dispersion relation of Alfv\'en waves propagating in a partially ionized plasma in the presence of an equilibrium flow of particles. Section \ref{S-figures} focuses on the numerical solutions of the dispersion relation and discusses the modifications in the damping rate of waves due to the presence of plasma flows for different ionisation degrees. Finally, in Section \ref{S-Conclusion} we present our conclusions, summarising our main findings and their implications for Alfv\'en wave propagation in partially ionised plasmas.
%---------------- 
  %{\bf --- Road map of the article} \\
    % When writing a article, it is worth to define a title for each paragraph:
    %   gives a roadmap for the article logic.
    % Replacing automatically all the ``%{\bf'' by ``{\bf'' 
    % brings these titles in the typeset of the article.

%%%%%%%%%%%%%%%%%%%%%%%%%%%%%%%%%%%%%%%
\section{Mathematical Background and the Dispersion Relation of Alfv\'en Waves}
\label{S-aug}

In a partially ionised solar plasma, we consider a frequency regime that is of the order of the collisional frequency between particles; therefore, we consider the dynamics within the framework of a two-fluid description where the charged particles (electrons and positive ions) are strongly coupled and they form a single charged fluid. Neutral atoms interact with charged particles through closed-range collisions, which is an effective mechanism for energy and momentum transfer between species. In our analysis, we are going to denote the charged and neutral species by indices {\it i} and {\it n}, respectively. The infinitely extended system is permeated by a homogeneous unidirectional magnetic field printed in the positive $z$ direction (${\bf B}_0=B_0{\bf \hat z})$, and waves are described within Cartesian geometry.  

We assume that the charged and neutral species have a steady and constant flow along the magnetic field, and these are denoted by $v_{0i}$ and $v_{0n}$, respectively. Since we will concentrate on Alfv\'en waves only, we will consider a pressure-less plasma in which the Alfv\'en waves are polarised in the $y$ direction and propagate along the background magnetic field. 
%The geometry of our problem is represented schematically by Fig (\ref{fig1}) . 
%\begin{figure}[h!]
%\centering
%\includegraphics[scale=0.25]{Geometry of the AW propagating in the direction of the mag netic field.jpg}
%\caption{Schematic representation of the geometry used in our problem. Alfv\'en waves are propagating along the $z$-direction and are polarised in the $y$-direction.}
%\label{fig1}
%\end{figure}\\
With these considerations, the linearised two-fluid MHD equations, therefore, are given by 
%%%%%%%%%%%%%%%%%%%%%%%%%%%%%%%%%
\begin{equation} 
\rho_{0i}\left( \frac{\partial }{\partial t}+{\bf v}_{0i}\cdot \nabla\right) \textbf{v}_i=\frac{1}{\mu }(\nabla
\times \textbf{b})\times \mathbf{B}-\rho_{0i}  \nu_{in} (\textbf{v}_i-\textbf{v}_n),
  \label{eq:1}
\end{equation}
\begin{equation}
\rho_{0n}\left(\frac{\partial}{\partial t}+{\bf v}_{0n}\cdot \nabla\right) \textbf{v}_n=-\rho_{0i}  \nu_{in}(\textbf{v}_n-\textbf{v}_i),
  \label{eq:2}
\end{equation}
\begin{equation} 
    \frac{\partial \textbf{b}}{\partial t}=\nabla \times (\textbf{v}_i\times \mathbf{B})+\nabla \times (\textbf{v}_{0i}\times \mathbf{b}).
      \label{eq:3}
\end{equation}
%%%%%%%%%%%%%%%%%%%%%%%%%%%%%%%%%
In the above equations ${\bf v}_i$ and ${\bf v}_n$ are the ion and neutral velocity perturbations, ${\bf B}$ and ${\bf b}$ are the equilibrium and perturbed magnetic fields, $\rho_{0i}$ and $\rho_{0n}$ are the equilibrium ion and neutral mass densities, $\mu$ is the magnetic permeability of free space, while $\nu_{in}$ and $\nu_{ni}$ are the ion-neutral and neutral-ion collisional frequencies. Since we consider elastic collisions between particles, we can write $\rho_{i} \nu_{in}$ = $\rho_{n} \nu_{ni}$. For the Alfv\'en waves studied here, the pressure gradients do not contribute in the leading order. For the sake of completeness, we provide quantitative estimates of these neglected terms in Appendix A that show that ion and neutral pressure forces are negligible compared with collisional drag across the regimes explored.

Moreover, the induction equation (\ref{eq:3}) should, strictly speaking, include a resistive term arising from electron collisions with both ions and neutrals. However, for the sake of analytical clarity, we have chosen to neglect this term. We emphasise that Cowling diffusion, while known to be significant in the upper chromosphere \citep{khodachenko2004}, is deliberately omitted from the present analysis to isolate the effects of ion-neutral collisional friction. Accordingly, our results are most applicable to chromospheric regions where collisional drag is the dominant dissipative process, such as the lower-to-middle chromosphere, where the collisional coupling between particles remains dominant.

Perturbations are Fourier-analysed and they are written proportional to the exponential factor $e^{i(kz-\omega t)}$, where $k$ is the real wavenumber and $\omega = \omega_r+i\omega_i$ is the complex frequency of waves, where the real part describes the propagation of waves and the imaginary part their damping rate. Since we are dealing with Alfv\'en waves polarised in the $y$-direction, we write ${\bf v}_{\alpha}=(0,v_{\alpha},0)$, where the index $\alpha=i,n$ denotes the two species of the problem and ${\bf b}=(0,b,0)$. 
%%%%%
%The system of equations is given as follows:

%1. For the ion momentum equation:
%\[
%\rho_{0i} \left( \frac{\partial v_{i}}{\partial t} + v_{0i} \frac{\partial v_{i}}{\partial z} \right) = \frac{1}{\mu} \left( \frac{\partial b}{\partial z} \right) B - \rho_{0i} \nu_{in} (v_{i} - v_{n})
%\]

%2. For the neutral momentum equation:
%\[
%\rho_{0n} \left( \frac{\partial v_{n}}{\partial t} + v_{0n} \frac{\partial v_{n}}{\partial z} \right) = - \rho_{0n} \nu_{ni} (v_{n} - v_{i})
%\]

%3. For the induction equation (magnetic field evolution):
%\[
%\frac{\partial \mathbf{b}}{\partial t} = \nabla \times (\mathbf{v}_i \times \mathbf{B}) +\nabla \times (\mathbf v_{0i} \times \mathbf{b})
%\]
%%%%%

Employing an extended, yet straightforward computation, the dispersion relation of waves travelling through partially ionised plasma in the presence of steady plasma flows can be written as
\[
\omega^3+ \left[i\nu_{in} (1+\frac{1}{\chi}) - k( 2 v_{0i}
+ v_{0n}) \right] \omega^2
\]
\[
+ \left\{k^2 (v_{0i}^2+2v_{0i}v_{0n} -c_A^2)-ik \nu_{in}\left[v_{0i}\left(\frac{2}{\chi}+1\right)+v_{0n}\right]\right\} \omega 
\]
\begin{equation}
+i k^2 \nu_{in} \left[v_{0i} v_{0n} +\frac{1}{\chi} (v_{0i}^2- c_A^2) \right] + k^3 v_{0n} (c_A^2 - v_{0i}^2) = 0,
 \label{eq:4}
\end{equation}
where $\chi=\rho_{0n}/\rho_{0i}$ is the ionization fraction of the plasma and $c_A^2=B_0^2/\mu\rho_{0i}$ is the square of the Alfv\'en speed.
  
\begin{figure} [!ht]   
\centering
\includegraphics[width=110mm]{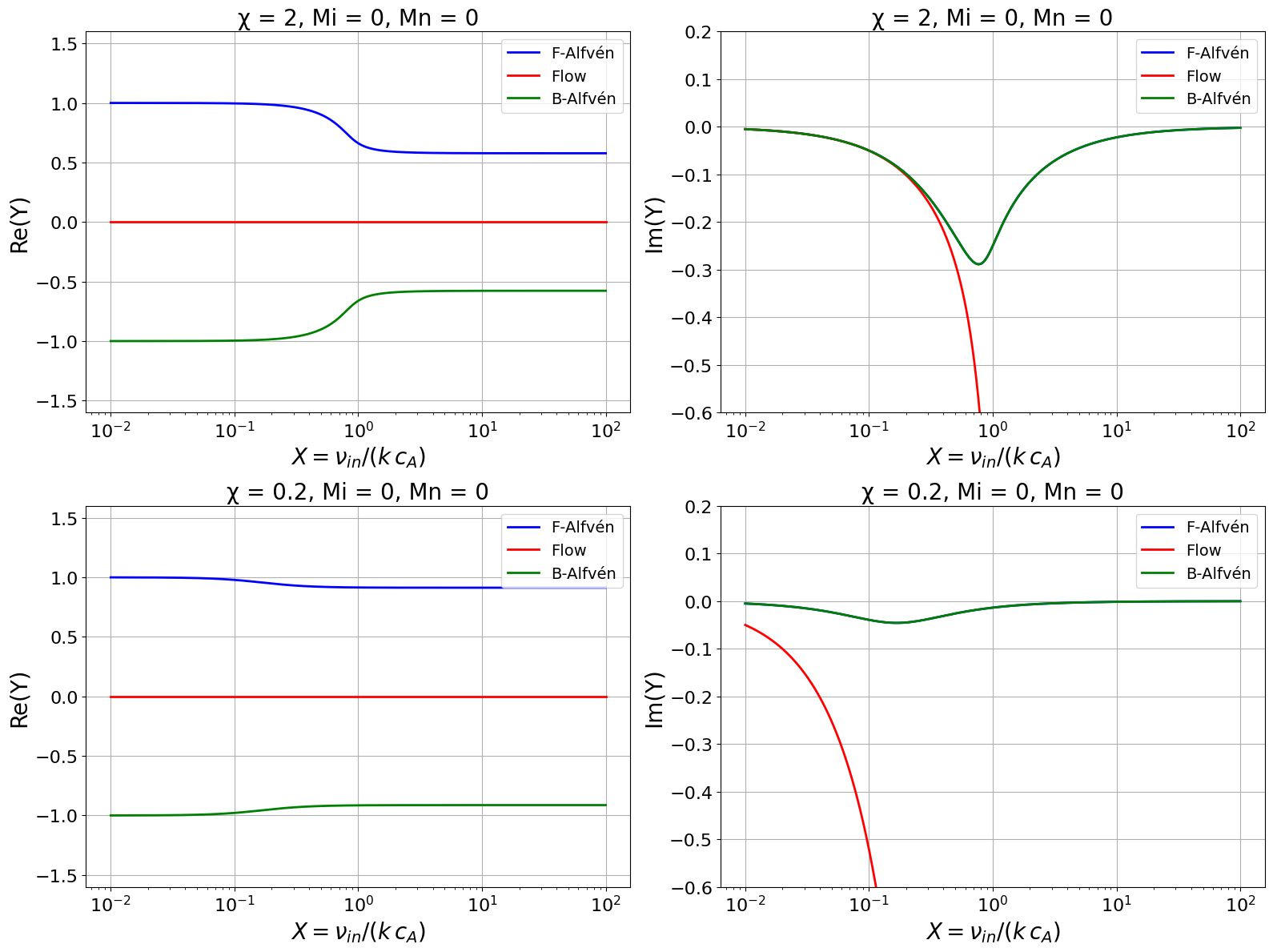}\small
\caption{The real (left column) and imaginary (right column) parts of $Y$ for weakly
ionised ($\chi =2$, top row) and strongly ionised ($\chi =0.2$, bottom row) plasmas in a static equilibrium $( M_{i}=M_{n}=0)$. The variation
of $Y$ is represented as a function of the dimensionless variable,
$X$. The forward (F) and backwards (B) propagating Alfv\'{e}n waves are shown
by green and blue lines, respectively. The mode shown in red is the non-propagating
(evanescent) entropy mode.}
\label{fig1}
\end{figure}

%%%%%%%%%%%%%%%%%%%%%%%%%%%%%%
To simplify our task, we write the above dispersion relation in dimensionless form and introduce the dimensionless parameters $X=\nu_{in}/k c_A$ denoting the strength of collisions, $Y=\omega/k c_A$ the dimensionless frequency of waves, $M_i=v_{0i}/c_A$ and $M_n=v_{0n}/c_A$ as the ion and neutral Mach numbers. In this form, the damping rate of waves is associated with the imaginary component of $Y$, whereas the real part of $Y$ is associated with the wave frequency. At the same time, the real and imaginary parts of $Y$ can also be interpreted as the waves' phase speed in units of Alfv\'en speed. In the new notations, the dispersion relation of Alfv\'en waves becomes 
\[
Y^3+ \left[iX \left(1+\frac{1}{\chi}\right)-2M_{i}- M_{n}\right] Y^2+
\]
\[
+\left\{M_{i}^2+2 M_{i} M_{n} -1 -iX\left[M_{i}\left(\frac{2}{\chi}+1\right)+ M_{n} \right]\right\} Y +
\]
\begin{equation}
+iX \left[M_{i} M_{n} +\frac{M_{i}^2- 1}{\chi}  \right] +  M_{n} (1 - M_{i}^2) = 0.   
\label{eq:10}
\end{equation}
If the two Mach numbers are set to zero, i.e., \( M_{0i}=M_{0n}=0\) (static equilibrium), the above dispersion relation becomes identical to the relation obtained earlier by \cite{soler2013alfven}. Figure \ref{fig1} shows the variation of the dimensionless frequency of waves with respect to the dimensionless collisional rate between particles.

%%%%%%%%%
These figures clearly show that in the absence of collisions (or very low collisional coupling between particles), Alfv\'en waves propagate with frequencies that are identical to their fully ionised plasma counterpart ($\omega=\pm kv_A$, or $Y=\pm 1$). With the increase of the collisional rate, the frequency of Alfv\'en waves decays (irrespective of the ionisation degree of the plasma) up to the level when the collisional frequency matches the natural frequency of Alfv\'en waves ($X\approx 1$), beyond which the plasma becomes a strongly coupled system. In this case, the frequency of Alfv\'en waves saturates, i.e. it becomes independent of the collisional rate between particles. In all cases, the frequency of waves is lower when the ion-neutral system is strongly coupled and the Alfv\'en waves transition from pure ion Alfv\'en modes to combined ion-neutral Alfv\'en waves with a reduced effective Alfv\'en speed $B_0/\sqrt{\mu_0(\rho_i+\rho_n)}$. Now the inertia of ions is increased by the presence of collisionally coupled neutrals. It is also clear that the modifications in the frequency of waves due to collisions are more pronounced in the case of weakly ionised plasmas. The imaginary part of the frequency describing the damping rate due to collisions reveals that waves will have their strongest damping when the collisional frequency approximately matches the natural frequency of the plasma, as this frequency regime proves the most ideal setup for an effective transfer of momentum between species and the damping rate of the forward and backwards Alfv\'en waves is identical. It is also clear that in the weakly and strongly coupled cases, Alfv\'en waves have weak damping. When the collisional frequency of waves becomes very large, the ion-neutral mixture behaves like a single fluid, and that explains the similar behaviour of wave damping in the two extreme cases. Alfv\'en waves propagating in a weakly ionised plasma undergo heavier damping than when propagating in a strongly ionised plasma.

A special class of solutions of the dispersion relation are denoted by the red line, and these are the modes that are non-propagating, very often labelled as "entropy" modes. In the context of Alfv\'en modes, this terminology is misleading as pure Alfv\'en waves do not generate entropy modes because they are incompressible and do not modify the density or pressure of the system. However, dissipative processes (like the collisions of particles considered in our study) can generate entropy modes in a different way. While ions are tied to the magnetic field, moving with the Alfv\'en waves, neutrals, not feeling the magnetic field directly, tend to lag behind. Collisions between the two species try to enforce coupling, but imperfect coupling leads to momentum exchange and energy dissipation. This dissipation means that wave energy is converted into entropy variations (localized heating), effectively generating entropy modes. In our analysis, we will use the terminology of "flow" mode to denote the modified "entropy" mode due to plasma flows and their connection to neutral advection in the presence of steady flow. From Figure \ref{fig1}, it is obvious that the real part of these modes is zero, while the imaginary part of the solutions shows very strong damping, and their damping rate increases significantly with the increase of the collisional rate between particles. 

The problem we are going to investigate can be simplified by considering a coordinate system attached to ions; therefore, using a Galilean transformation of the coordinate system, we can introduce the new quantities $Y^\prime=Y-M_i$, $M_i=0$ and $M_n^\prime=M_n-M_i$. As a result, the dispersion relation \ref{eq:10} transforms into
\begin{equation}
Y'^3 
+ \left[ iX\left(1 + \frac{1}{\chi}\right) - M_n' \right] Y'^2 
- \left[ 1 + iX M_n' \right] Y' 
+ \left[ M_n'-\frac{iX}{\chi} \right] = 0,
\label{eq:10.1}
\end{equation}
where now $M_n'$ denotes the relative drift between neutrals and ions and $Y'$ is now measured in the ion rest frame. For simplicity, in what follows,  we are going to drop the dash symbol.

%-##########################################
\section{Numerical Solutions of the Dispersion Relation} %%%%%%%%%%%%%%
\label{S-figures}
  %{\bf --- Main features} \\

To investigate the influence of background plasma flows on the properties of Alfv\'en waves, we evaluate numerically the roots of the dispersion relation (\ref{eq:10.1}) across a range of ion and neutral flow conditions. Our study examines two distinct cases, systematically varying the Mach numbers of constituent species, while maintaining different ionisation levels ($\chi=2$ and $\chi=0.2$), corresponding to weakly and strongly ionized plasmas, respectively.

%#####################################figure 2
%\begin{figure}
%    \centering
 %     \begin{subfigure}[b]
   %    \centering
%            \includegraphics[width=60mm]{R1_chi2_Mn_varying.png}
 %        \centering
 %        \includegraphics[width=60mm]{Im1_chi2_Mn_varying.png}
 %                \caption{The variation of the real and imaginary parts of the dimensionless variable $Y$ as a function of the dimensionless collisional variable $X$, with $\chi=2$ and $M_i=0.1$, for varying $M_n$. The two panels show the real and imaginary parts of $Y$. The forward (F), backwards (B) Alfv\'en waves, together with the flow mode, are shown by different colours, with $X$ plotted on a logarithmic scale. The black line denotes the boundary between forward and backwards propagating flow modes.}
 %    \end{subfigure}
% \label{fig2}
%\end{figure}
%#############################################
%#####################################figure 2.1
\begin{figure}
    \centering
      \begin{subfigure}[b]
       \centering
            \includegraphics[width=60mm]{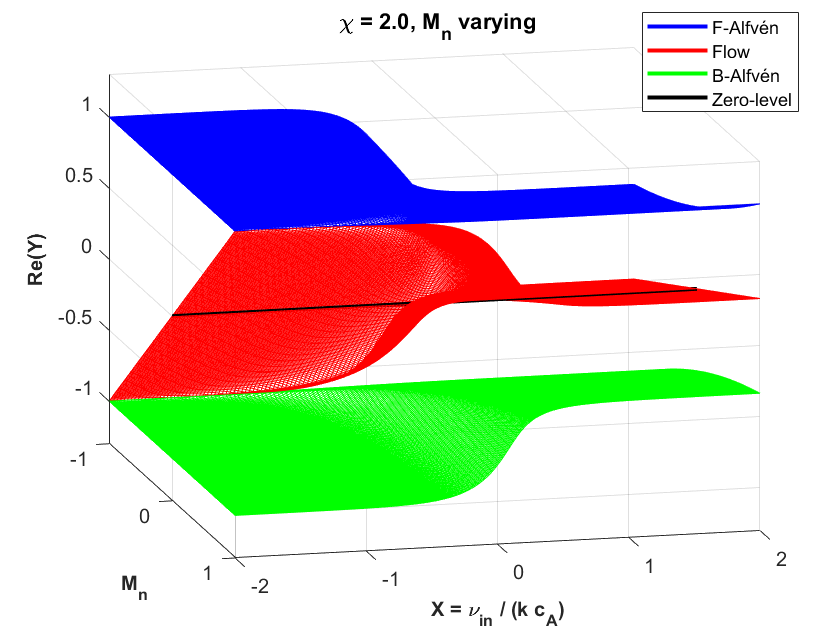}
         \centering
         \includegraphics[width=60mm]{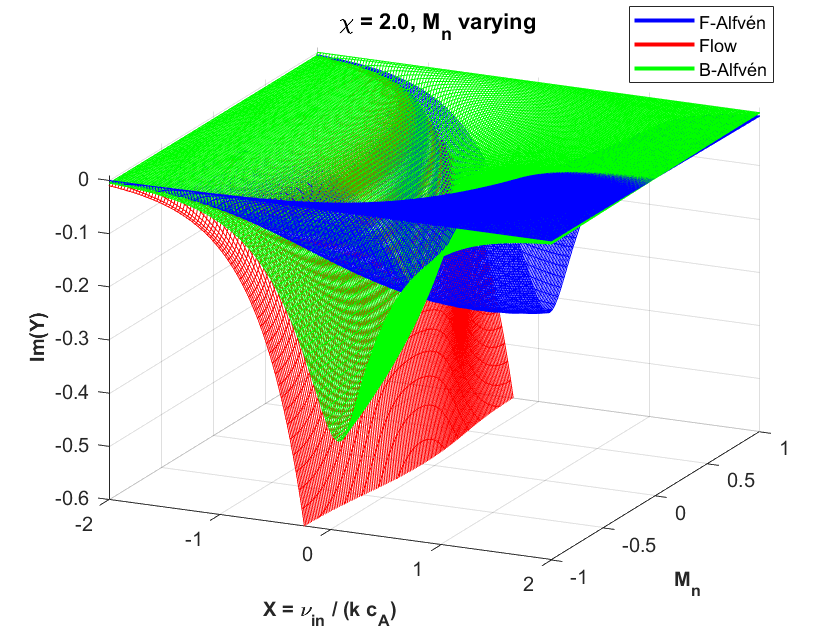}
                 \caption{The variation of the real and imaginary parts of the dimensionless variable $Y$ as a function of the dimensionless collisional variable $X$, with $\chi=2$ for varying $M_n$. The two panels show the real and imaginary parts of $Y$. The forward (F), backwards (B) Alfv\'en waves, together with the flow mode, are shown by different colours, with $X$ plotted on a logarithmic scale. The black line denotes the boundary between forward and backwards propagating flow modes.}
     \end{subfigure}
 \label{fig2}
\end{figure}
%#############################################

%\subsection{Weakly ionised plasmas ($\chi=2$), $M_i=0.1$}
 \subsection{Weakly Ionised Plasmas (\texorpdfstring{$\chi=2$}{chi=2})} %\texorpdfstring{$M_i=0.1$}{Mi=0.1}

First, we investigate the behaviour of Alfv\'en waves in a partially ionised plasma relevant to deeper layers of the solar atmosphere, where we consider that the number density of neutrals is twice that of ions ($\chi=2$), i.e. we deal with a weakly ionized case. The first set of results investigates the behaviour of Alfv\'en waves when the relative speed of neutrals, $M_n$, is varied between $1$ and $-1$ to assess the influence of the flow of neutral species on wave propagation and damping. The negative values of the neutral Mach number denote counter-flowing neutral species compared to the direction of the equilibrium magnetic field. The real and imaginary parts of the solutions are shown in the two panels of Figure 2. To better understand the changes in the damping rate of waves, we also provide two cuts at $M_n=\pm 0.5$ in Figure 3.

In the left-hand side panel of Figure 2 displaying the real part of solutions, the forward (blue surface) and backwards (green surface) Alfv\'en waves are propagating in a symmetric way. In dimensionless units, in the absence of collisions ($X=0$), the two Alfv\'en modes propagate with a dimensionless frequency $Y=\pm 1$. The dependence of the dimensionless frequency of waves with respect to the variable $X$ resembles the variation shown in the top right panel of Figure \ref{fig1}. Therefore, with the increase of the rate of collisional coupling between ions and neutrals, the inertia of field lines increases, leading to a smaller propagation speed (or frequency) of Alfv\'en waves. When the collisional rate is high, the ion-neutral mixture behaves like a fluid, and the propagation speed of the two Alfv\'en modes becomes the effective Alfv\'en speed. 
%================================================
%\begin{figure}
%    \centering
 %     \begin{subfigure}[b]
%       \centering
 %           \includegraphics[width=60mm]{Im2D_chi2_Mi0p1_Mn0p5.png}
 %        \centering
%         \includegraphics[width=60mm]{Im2D_chi=2_Mi=0.1_Mn=-0.5.png}
%                 \caption{The variation of the imaginary parts of the dimensionless variable $Y$ as a function of the dimensionless collisional variable $X$, with $\chi=2$ and $M_i=0.1$, for the particular values of $M_n=\pm 0.5$.}
 %    \end{subfigure}
% \label{fig:2.1}
%\end{figure}
%========================================
%================================================
\begin{figure}
    \centering
      \begin{subfigure}[b]
       \centering
            \includegraphics[width=60mm]{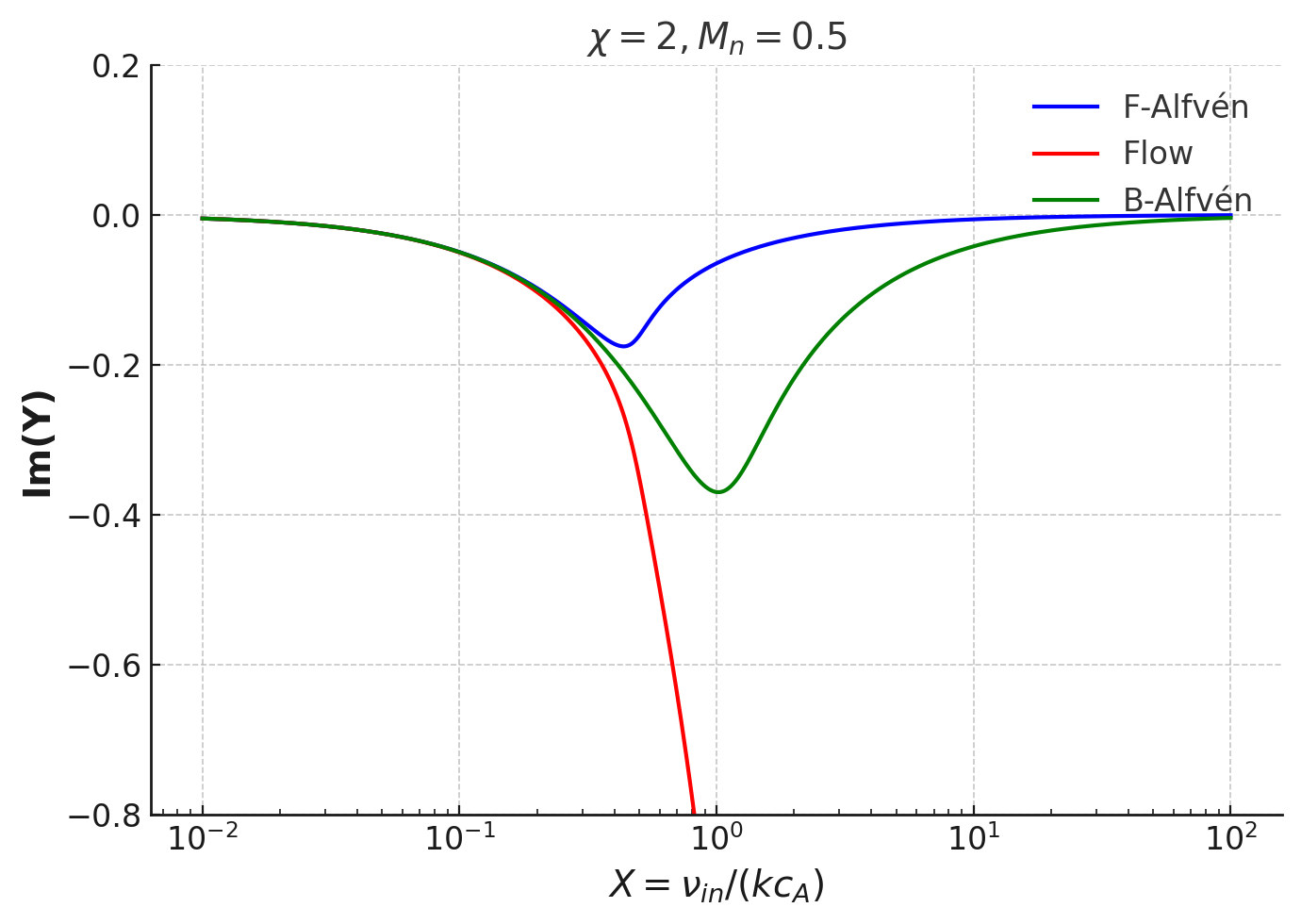}
 %        \centering
         \includegraphics[width=60mm]{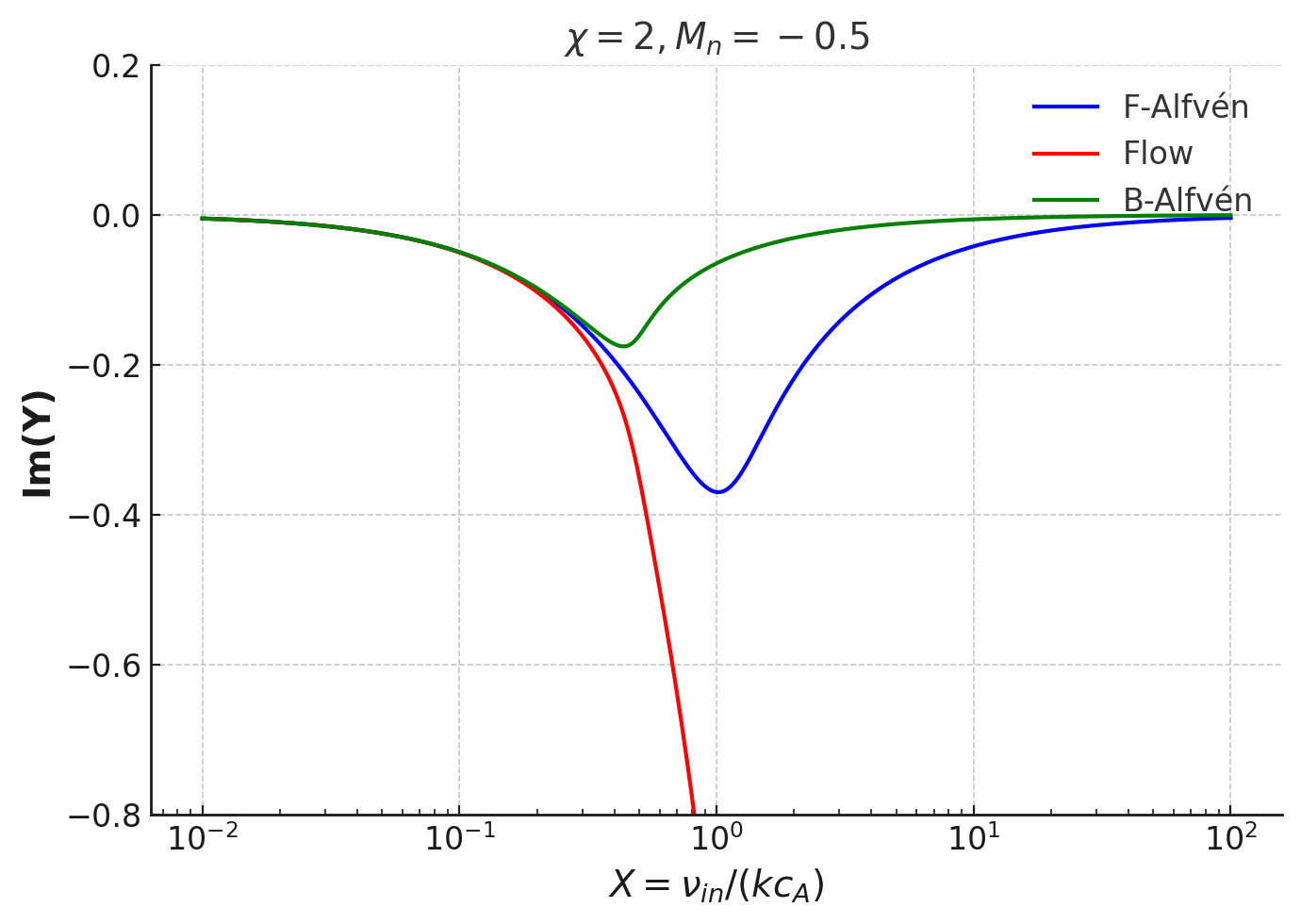}
                 \caption{The variation of the imaginary parts of the dimensionless variable $Y$ as a function of the dimensionless collisional variable $X$, with $\chi=2$ for the particular values of $M_n=\pm 0.5$.}
     \end{subfigure}
 \label{fig:2.1}
\end{figure}
%========================================
One fundamentally different aspect of the investigated problem compared to the case of Alfv\'en waves in a static equilibrium is the appearance of a third mode, called the flow mode (red surface). This mode propagates solely due to the presence of ambient neutral flow. As a result, the entropy (non-propagating) mode that was present in the static plasma becomes advected by the flow and becomes propagating. This mode is not a natural wave as it has no restoring force, and it propagates with the speed of the neutral flow, $M_n$, in the absence of collisions.

For a weak degree of collisionality, the neutrals are weakly coupled to ions, and the flow mode varies linearly with $M_n$. With the increase in the collisional rate, the flow mode displays an influence due to the presence of magnetised ions, and this behaviour can be easily understood. When $X\approx 1$, ion-neutral collisions become frequent enough so that neutrals try to drag ions along, and vice versa. Since ions are tied to the magnetic field, neutrals feel an indirect magnetic restoring force through their coupling to the ions. As a result, the neutral flow mode develops magnetically influenced character and the frequency of the flow mode approaches the frequency of the nearby Alfv\'en mode (approaches the forward Alfv\'en mode for $M_n>0$ and backwards Alfv\'en mode when $M_n<0$). The bending of the flow mode reflects partial hybridisation of the flow mode. Although modes undergo changes in their frequency for intermediate collisional rates, they maintain their identity, i.e. no mode conversion takes place. 

The variation of the imaginary part of $Y$ in terms of the dimensionless collisional frequency, $X$, and the neutral Mach number (right-hand side panel of Figure 2) reveals that for small values of the collisional coupling between species, the three waves propagate with very low damping. When $X\gg 1$, the particles are tightly coupled and the mixture behaves like a single fluid. In this regime, due to the strong coupling of neutrals to ions, the relative drift of the two species is very much reduced, leading to a minimal damping of Alfv\'en waves. The two panels in Figure 3 reveal that at an intermediate level of collisions (near $X=1$), the damping of the two Alfv\'en modes shows an asymmetry driven by the coupling with the flowing neutral species. Accordingly, for neutrals flowing in the direction of the magnetic field, the neutrals coupled to the forward-propagating Alfv\'en wave reduce the damping rate of these waves, while the backwards-propagating wave suffers more friction. A symmetrical phenomenon occurs for neutrals flowing in the negative direction; now the forward propagating Alfv\'en waves have stronger damping, while the presence of neutrals decreases the damping rate of backwards propagating waves.  These results confirm directional damping asymmetry due to neutral counter-flow. In addition, the presence of flowing neutrals also moves the peak of the damping rates towards smaller values of $X$, suggesting that the flow of neutrals enhances the momentum exchange between species.

%The damping rates of the two modes attain their maximum in this region, therefore collisions are most effective when teh collisional frequency of particles matches the natural frequency of the Alfv\'en waves. This region is also where the momentum exchange between species is the most effective. 

The imaginary part of the flow mode has an interesting behaviour. After a very weak damping (in line with the other two Alfv\'en waves) for weakly coupled plasma, the flow mode quickly undergoes a heavy damping, similar to the result obtained in the case of a static background. In this figure (like for the remaining figures), we limit the extent of the vertical axis to be able to focus on the damping rates of the two Alfv\'en waves; however, the damping rate of the flow modes keeps increasing, so these waves are overdamped. Around $X=1$, ions and neutrals collide frequently enough to interact, but not so frequently that they move together. This regime corresponds to a strong momentum exchange in which neutrals are able to drag the ions, while ions respond via the magnetic tension in the field. The collisional coupling between species leads to a dissipative force that acts upon neutrals. As specified earlier, unlike Alfv\'en modes, the flow mode has no inherent restoring force and it relies entirely on fluid inertia and collisions. When collisions become important, there is nothing to balance the frictional loss; therefore, flow modes experience heavy damping. For a collision-dominated plasma ($X \gg 1$), the species are effectively coupled together, leading to vanishing drift of species, i.e. vanishing friction. As a result, the flow mode fades away and the only mode that survives is the Alfv\'en mode, which propagates with the effective Alfv\'en speed.

\begin{figure}
    \centering
       \begin{subfigure}[b]
       \centering
         \includegraphics[width=60mm]{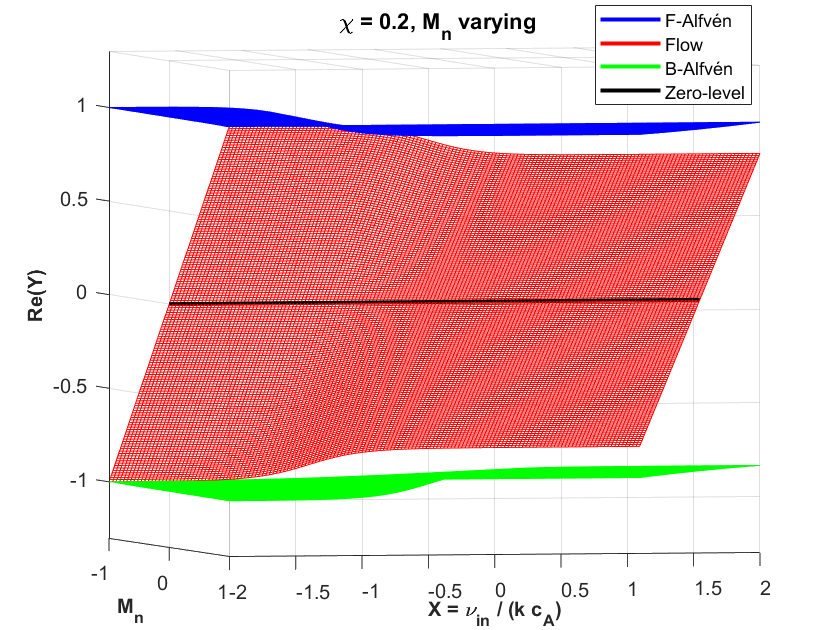}
 %        \centering
         \includegraphics[width=60mm]{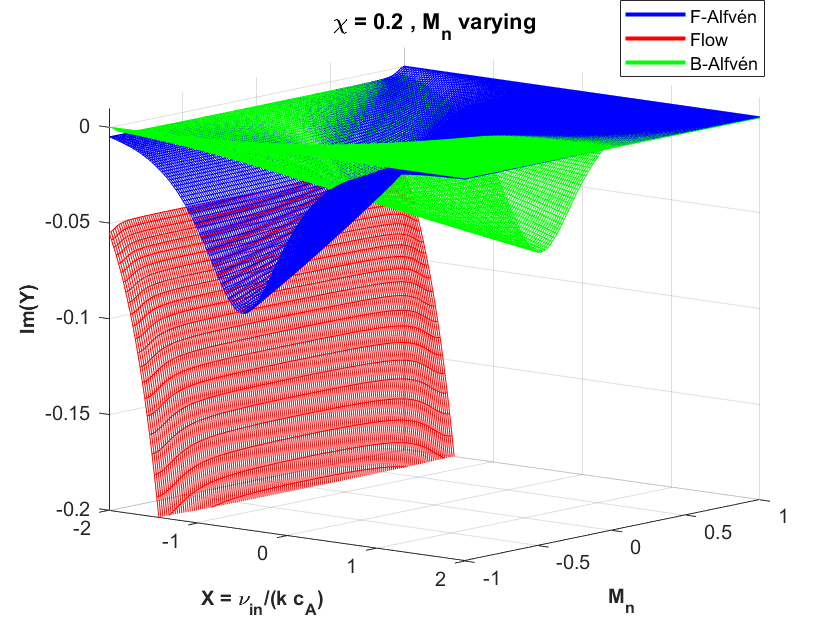}
              \caption{Similar to Figure 2, but here the curves are obtained for $\chi=0.2$ and for varying $M_n$. } 
     \end{subfigure}
    \label{fig4}
\end{figure}

%======================

%\subsection{Strongly ionised plasma ($\chi=0.2$), $M_i=0.1$}

\subsection{Strongly Ionised Plasma (\texorpdfstring{$\chi=0.2$}{chi=0.2})} %\texorpdfstring{$M_i=0.1$}{Mi=0.1})}

Figure (4) shows the variation of the dimensionless frequency of waves in terms of the dimensionless collisional frequency and the relative neutral Mach number when the plasma is strongly ionised ($\chi=0.2$). 
%To better understand the intricate interaction between modes, we also present a 2D cut through the surfaces at $M_n=\pm 0.5$ (see Fig. 6).

First of all, compared to the weakly ionised case (see Figure 2), the forward and backwards Alfv\'en branches are largely flat, with only slight shifts in frequency as $M_n$ varies. This result is easy to understand as in the case of strong ionisation ($\chi\ll 1$), ions dominate the inertia of the system. Neutrals play a minor role in the total momentum, and their drift (positive or negative) produces small Doppler shifts through collisional coupling, but not strong frequency modifications. That is why the flow mode varies almost linearly with the strength of the neutral flow and couples very weakly to the Alfv\'en modes for large values of $X$. Thanks to this weak coupling, no mode conversion or avoided crossing is observed. Therefore, in the strongly ionised case, the flow mode behaves mostly like a hydrodynamic wave.

%#############################################
%\begin{figure} [h!]   
%\centering
%\includegraphics[scale=0.45]{Re--_2D--_chi0.2,Mi=0.1,M_n=0.5,-0.5.png}
%\caption{The real  parts of $Y$ for strongly ionised ($\chi=0.2$) plasma, $ M_i=0.1 $, $M_n=0.5$ (left panel) and $M_n=-0.5$ (right panel). The meaning of the colours used in this plot is identical to the ones used earlier.}
%\label{fig4.1}
%\end{figure}

The effects of the coupling between particles and flows can also be seen in the variation of the imaginary part of the dimensionless frequency displayed in the right-hand side panel of Figure 4, together with the 2D cuts through the surfaces at $M_n=\pm 0.5$ (Figure 5). The flow mode shows an increased damping with collisions despite the coupling with ions. However, this behaviour can be understood in the light of the hydrodynamic nature of the flow modes. 

Since we are dealing with a strongly ionised plasma, the Alfv\'en modes are mildly affected by the presence of flowing neutrals. At small collisional coupling, ions move freely and drag neutrals weakly, so the wave damping is small. The attenuation of wave peaks in the region near $X=1$, where friction between particles takes its maximal value. In the strongly collisional limit, $X\gg 1$, ions and neutrals are tightly coupled, reducing relative drift and thus damping. The asymmetry that can be seen in the 2D cuts between forward/backwards modes is due to the direction of neutral flow. In the presence of flowing neutrals, these introduce a directional drag on the plasma through the ion-neutral collisions. This drag is not symmetric with respect to the wave propagation direction. For neutrals propagating in the positive direction, the forward propagating Alfv\'en wave experiences a larger relative velocity with respect to the neutrals, leading to enhanced momentum exchange between particles, increasing collisional damping. In this case, the backwards Alfv\'en wave moves against the neutral flow, so the relative velocity is smaller, meaning less damping. When neutrals flow in the negative direction, the backwards wave experiences a larger relative velocity with respect to neutrals, leading to stronger damping. The forward wave moves into the neutral stream, reducing relative velocity and resulting in weaker damping. 

Compared with the case of weakly ionised plasmas, which proved to be a highly dissipative environment, especially when $X\sim 1$ (making them ideal for energy conversion and heating), strongly ionised plasmas support longer-lived Alfv\'en modes, which retain their structure and can propagate further. In weakly ionised plasmas, the magnetic field exerts less control on the bulk plasma motion, so neutral drag significantly enhances damping. In contrast, in strongly ionised plasmas, the dynamics is dominated by ions that confer an effective shield to the magnetic waves from neutral friction, leading to reduced damping. Weak ionisation introduces asymmetric damping for forward and backwards propagating Alfv\'en waves (clearly seen in the 2D cuts), because of directional effects induced by neutral flow. In contrast, in the strong ionisation limit, the damping profiles maintain nearly symmetric and mild damping, with small shifts in the damping profile. In weakly ionised plasmas, the neutral drag dominates, so the asymmetry in damping is pronounced. That is why the neutral flow breaks the symmetry between forward and backwards wave propagation. In contrast, in strongly ionised plasma, the plasma is tightly coupled to the magnetic field, so neutral flow has less impact, and damping becomes more symmetric.

%However, an interesting feature occurs around $X\approx 1$, where the real frequencies undergo avoided crossings. Now the imaginary parts show strong peaks in damping. This reflects that mode conversion is energetically lossy, i.e. the coupled modes are not ideal oscillations, but instead transfer energy to each other via collisions. This aspect highlights that avoided crossings are not just frequency shifts but involve energy redistribution and dissipation. The modes undergoing conversion absorb or share damping from their coupled counterpart. The strongest damping of the flow mode coincides with its transition into another wave mode, reinforcing the idea that energy is transferred between modes at this point. As collisional effects become dominant at higher $X$ values, the flow mode starts to disappear, and the plasma begins to behave as a single-fluid system where both species (ions and neutrals) move collectively. For positive values of $M_n$ damping is strongest in the backward propagating Alfvén mode.
%======================================
%\begin{figure} [!ht]   
%\centering
%\includegraphics[width=110mm]{Mi0p1_Mn_pm0p5.png}
%\caption{The imaginary parts of $Y$ for strongly ionised ($\chi=0.2$), $M_i=0.1 $, $M_n=0.5$ (left panel) and $M_n=-0.5$ (right panel). The colour coding for the modes is consistent with that used in the real part plots.}
%\label{fig4.2}
%\end{figure}
%====================================
%================================================
\begin{figure}
    \centering
      \begin{subfigure}[b]
       \centering
            \includegraphics[width=60mm]{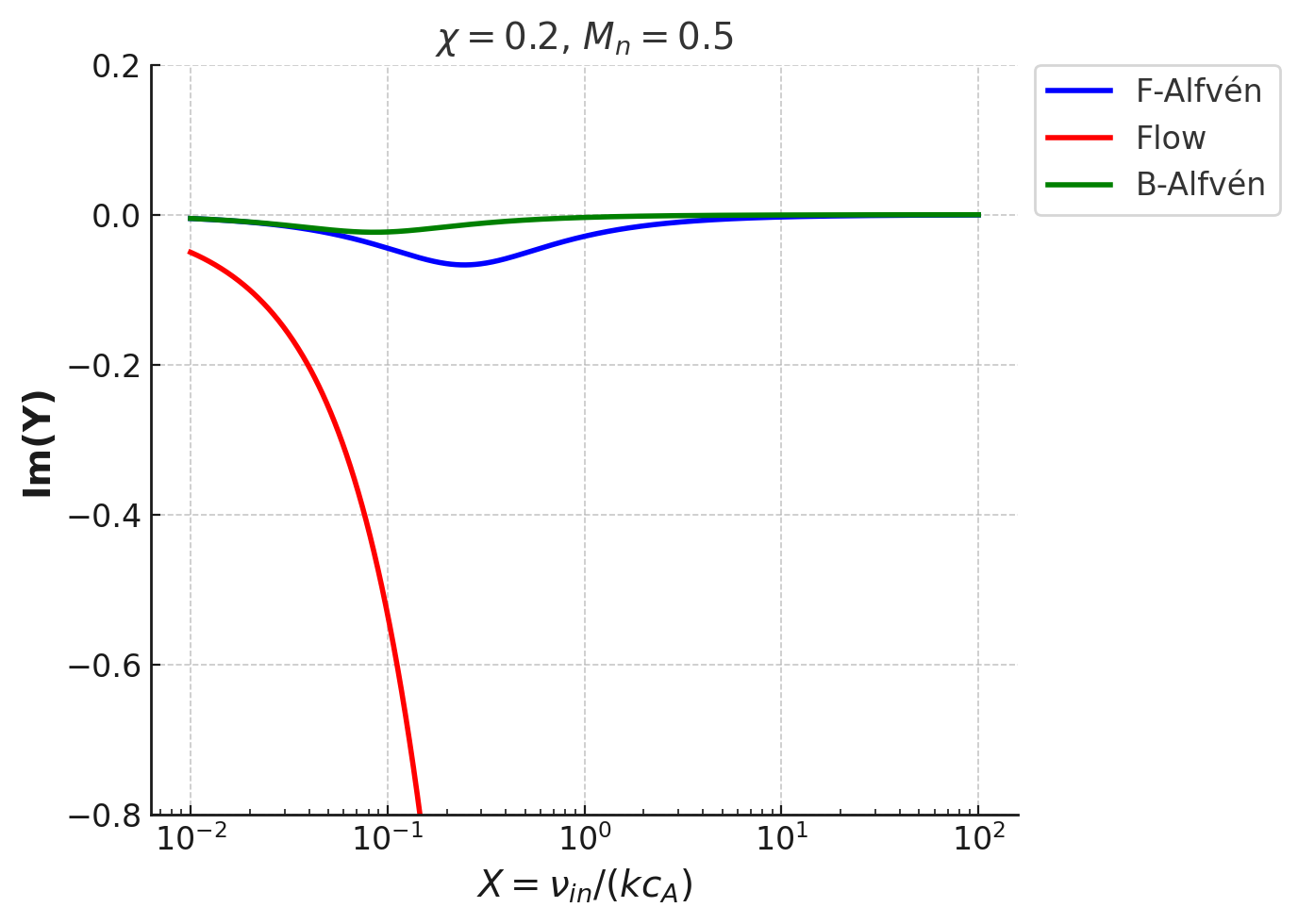}
 %        \centering
         \includegraphics[width=60mm]{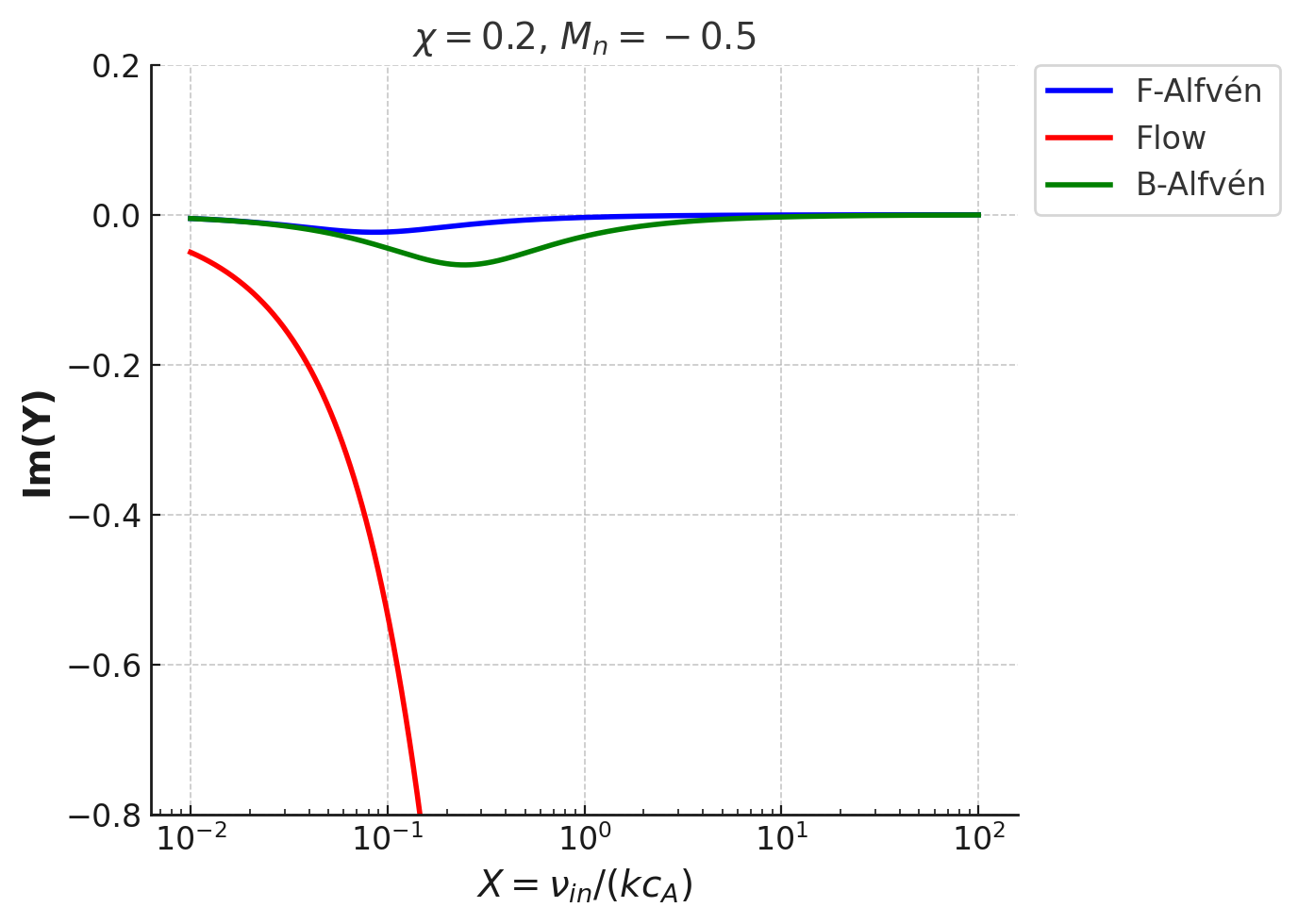}
                 \caption{The imaginary parts of $Y$ for strongly ionised plasma ($\chi=0.2$), when $M_n=0.5$ (left panel) and $M_n=-0.5$ (right panel). The colour coding for the modes is consistent with that used in the real part plots.}
     \end{subfigure}
 \label{fig:4.2}
\end{figure}
\begin{figure} [!ht]   
\centering
\includegraphics[width=120mm]{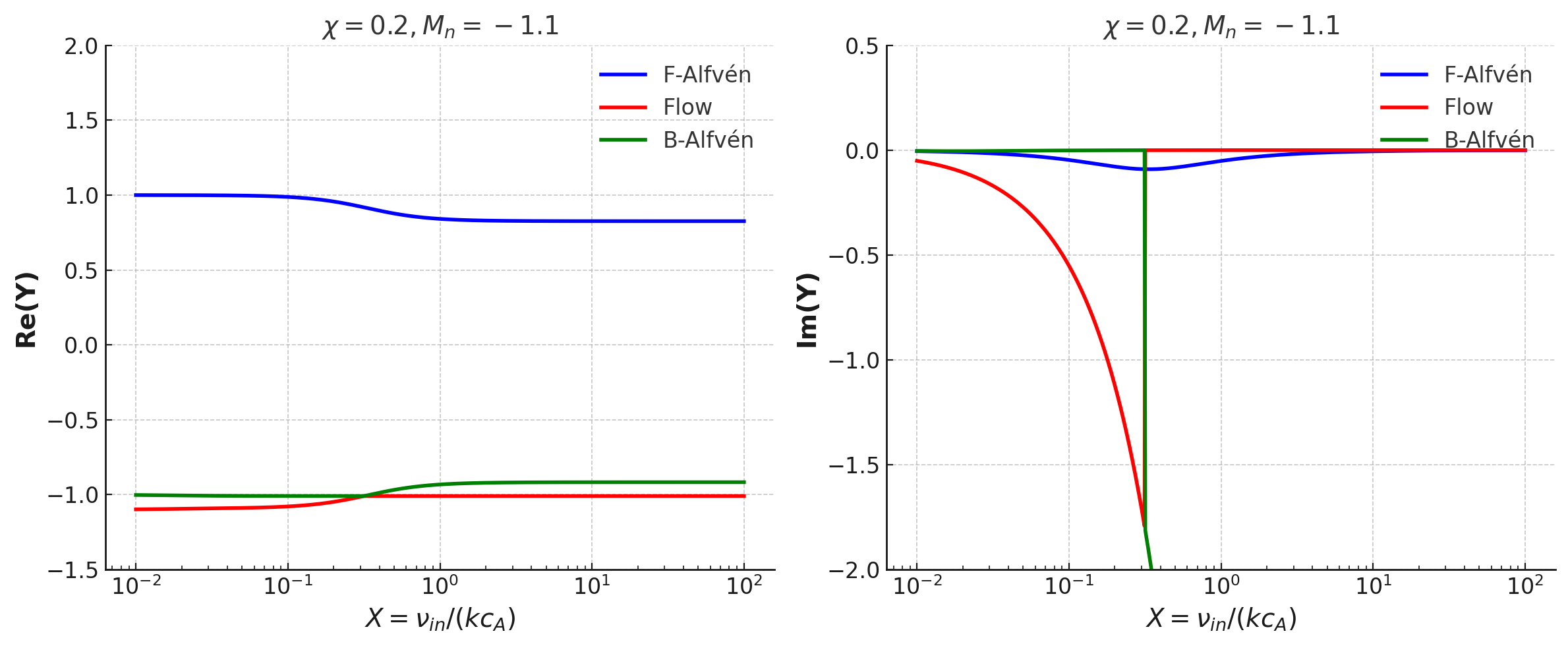}
\caption{Special case of mode conversion for $\chi =0.2$, and $M_n=-1.1$. The two panels show the real (left-hand) and imaginary (right-hand) parts of $Y$, highlighting mode conversion and damping.}
\label{fig6}
\end{figure}

%######################
\subsection{Special Case: Mode Conversion}

Before concluding our results, let us discuss a special case that illustrates the intricate role of the ion and neutral flows on the properties of waves. Let us consider a particular case of strongly ionised plasmas shown in Figure \ref{fig6}. In this case, Figure \ref{fig6} shows the variation of the real and imaginary parts of the dimensionless quantity $Y$ with respect to the collisional parameter $X$ when $M_n=-1.1$, i.e., we study the properties of the possible waves in the presence of a relatively strong counter-flowing species.  

In this case, the flow mode propagates backwards and at about $X\approx 0.3$ the flow and the backwards Alfv\'en mode merge, undergoing a mode conversion. In the strongly ionised case, ions dominate the dynamics; therefore, even weak collisions are enough to drag neutrals along, enhancing momentum exchange between ion and neutral fluids. The counter-flowing species enhance the coupling between modes. The flow mode, which is primarily dominated by the presence of neutrals at low values of collisional coupling, starts feeling the presence of magnetised ions through collisions. As a result, the frequency of the flow mode reduces, and at the conversion point it matches the frequency of the backwards propagating Alfv\'en mode. At conversion, the momentum and energy partition of the two modes match, and they resonantly exchange identities. During this process, the flow mode gains magnetic behaviour, and it becomes the new backwards Alfvén mode. On the other hand, the original backwards propagating Alfv\'en wave transitions to a neutral-damped flow-like mode. The process of mode conversion comes together with an exchange of energy between modes. While Alfv\'en modes carry primarily magnetic energy, flow modes are predominantly hydrodynamic waves, and during the conversion, these energy forms are interchanged between the interacting modes. 

The mode conversion is also visible in the variation of the damping rate of the three modes with the collisional coupling of species. The forward propagating Alfv\'en waves have a typical damping rate, similar to the cases discussed earlier. The damping rate of these modes reaches its maximum at a collisional rate where the exchange of momentum between species is optimal, i.e. near $X\approx 1$. On the other hand, the damping of the flow mode and the backwards propagating Alfv\'en wave is different. Flow modes have a very rapid damping, similar to the earlier results up to the point of conversion. Parallel to the interchange between the identities of flow and backwards Alfv\'en modes, the damping rate of the modes is also swapped. As a result, the flow mode becomes practically undamped, while the backwards Alfv\'en wave damps very effectively.

\section{Conclusion} %%%%%%%%%%%%%%%%%%%%%%%%%%%%%%%%%%%%%%%%
\label{S-Conclusion}

In this study, we investigated the propagation of Alfv\'en waves in partially ionised solar plasmas with field-aligned steady flows of charged and neutral components. Using a two-fluid model, we explored the modifications in the properties of Alfv\'en waves (in a reference system of reference attached to ions) induced by the presence of background flows and how the damping of these modes is changed. Our work constitutes an extension of previous studies of Alfv\'en waves in static plasmas by explicitly incorporating species-specific background flows.

By linearising the governing equations and applying a standard Fourier analysis, we derived the dispersion relation that describes the propagation and damping characteristics of waves. Our results indicate that the introduction of steady flows fundamentally modifies wave behaviour. The dispersion relation develops asymmetries, frequency shifts, and in some regimes, flow-induced mode coupling. The presence of neutral flow leads to the emergence of a new propagating mode (flow mode) that originates from the non-propagating entropy solution in the static case. The flow mode is associated with the advection of perturbations by the neutral flow. While not a true wave in the traditional sense (no restoring force), it acquires apparent propagation due to flow advection. This mode interacts with the Alfv\'en modes when collisional coupling becomes significant. The flow mode exhibits a strong damping, provided modes are not coupled. The flow mode facilitates interactions between modes, and it is a mode that has (for low levels of collisional coupling) hydrodynamic character. 

While ion-neutral collisions provide the mechanism for coupling, the neutral flow adds directionality, asymmetry, and energy into the system. The magnitude and direction of the neutral flow select which Alfv\'en mode the flow mode couples with more strongly. Relative flow between ions and neutrals creates a net drift, enhancing or reducing the effectiveness of momentum exchange. For $M_n>0$, the flow mode couples more with the forward Alfv\'en mode. In contrast, for $M_n<0$, the flow mode couples more with the backwards Alfv\'en mode. In other words, the flow of the neutral species can control the degree of mixing of the flow mode with one of the Alfv\'en waves. The same flow, together with particle collisions, leads to the hybridisation of modes and a distortion of the modes' identity. When the ion-neutral coupling is strong (intermediate $X$), the flow mode inherits magnetic characteristics, and Alfv\'en modes become "dragged" by the flow. As a result, we no longer have a clear separation between "magnetic" and "neutral" modes. All three modes become hybrids, but the degree of hybridisation depends on the relative flow speed of neutrals. 

In weakly ionised plasmas, neutral flows strongly influence the damping and phase speed of waves. The direction of neutral flow introduces asymmetries in the damping of forward and backwards propagating Alfv\'en waves, with enhanced dissipation occurring when wave propagation is counter to neutral drift.

In contrast, in strongly ionised plasmas, where ions dominate inertia, the effect of neutral flows is somehow smaller. However, under specific conditions, collisional coupling enables efficient energy exchange between the flow mode and a backwards-propagating Alfvén wave. This manifests as a clear mode conversion, characterised by the exchange of modal identity and energy content. The flow-induced mode conversion occurs most efficiently when the real frequencies of the two modes approach, and is accompanied by a complementary transfer of magnetic and kinetic energy, as demonstrated through energy partition analysis. During this conversion, the wave that was originally magnetically dominated becomes increasingly hydrodynamic in character, and vice versa. Our results suggest that even modest flows can strongly influence wave energetics and damping behaviour. 

The obtained results might form the theoretical background for the diagnostic potential of damping asymmetries in identifying the presence and direction of neutral flows in the solar chromosphere, where the differential flow between species is likely to occur.

Finally, our study used a number of simplifications that allowed us to derive a relatively simple dispersion relation. One critical assumption was the presence of a homogeneous flow. In reality, flows are often sheared, which gives rise to a more effective energy exchange between the background flow and waves, as well as energy exchange between different waves. Although these aspects were somehow recovered by our analysis, the presence of a shear flow can enhance these phenomena. It is our intention to expand the current analysis to shear flows in the near future. {In addition, our analysis focused solely on the simplest Alfv\'en waves to shed light on the effects introduced by homogeneous flows. We intend to expand the current analysis to other kinds of waves that could propagate in partially ionised solar plasmas.}  We deliberately considered collisional coupling between particles as the dominant transport mechanism. As specified earlier, in the upper chromosphere, resistive effects due to the Cowling resistivity might become dominant, in which case additional damping mechanisms have to be added to the governing equation that will lead to additional damping of Alfv\'en waves.

\begin{acks}
 NFA acknowledges the King Khalid University and the Ministry of Education in the Kingdom of Saudi Arabia for their financial support. IB, VF, and GV are grateful to the Science and Technology Facilities Council (STFC) grants ST/V000977/1, ST/Y001532/1. VF, GV thank the Institute for Space-Earth Environmental Research (ISEE, International Joint Research Program, Nagoya University, Japan). VF, GV, IB thank the Royal Society, International Exchanges Scheme, collaborations with Instituto de Astrofisica de Canarias, Spain (IES/R2/212183), Institute for Astronomy, Astrophysics, Space Applications and Remote Sensing, National Observatory of Athens, Greece (IES/R1/221095), Indian Institute of Astrophysics, India (IES/R1/211123) and collaboration with Ukraine (IES/R1/211177) for the support provided. 
\end{acks}
 \begin{ethics}
\begin{conflict}
 The authors declare that they have no conflicts of interest.
\end{conflict}
 \end{ethics}

\appendix
\section{Quantitative Estimates for Pressure Forces in the Momentum Equations \ref{eq:1} and \ref{eq:2}}

The pressure gradient terms in Equations \ref{eq:1}--\ref{eq:2} were neglected only on the basis that in a fully ionised plasma, the pressure gradient does not contribute to the momentum balance as Alfv\'en waves are incompressible waves. In this Appendix, we show that this statement remains true even in a partially ionised plasma considered by our study. Let us first consider the momentum equation for ions. Taking into account all the forces that might act upon a plasma element of unit volume, we can write that 
\begin{equation} 
\rho_{0i}\left( \frac{\partial }{\partial t}+{\bf v}_{0i}\cdot \nabla\right) \textbf{v}_i=\frac{1}{\mu }(\nabla
\times \textbf{b})\times \mathbf{B}-\rho_{0i}  \nu_{in} (\textbf{v}_i-\textbf{v}_n)-\nabla p_i,
  \label{eq:A1}
\end{equation}
where $p_i$ is the ion pressure perturbation and all the other variables have been defined earlier. For the Alfv\'en waves studied in the present paper (velocity polarized in $y$ direction, propagation along the $z$ direction), the perturbations are incompressible to leading order, so $\nabla \cdot {\bf v}_i=0$ and $\rho_i=0$, hence $\nabla p_i=0$ at leading order. To be conservative, we bound the largest possible pressure-gradient contribution by allowing a small residual compressibility, so that $\nabla \cdot {\bf v}_i\neq 0$. In order to cast the residual compressibility of the system, we introduce the small parameter $0<\epsilon_i\ll 1$ defined as
\[
\epsilon_i=\frac{v_{i\parallel}}{v_{i}}.
\]
Using the standard mass conservation of ions, the same small parameter $\epsilon_i$ can also be written as
\[
\epsilon_i=\frac{\omega}{kv_i}\frac{\rho_i}{\rho_{0i}},
\]
so $\epsilon_i$ also measures the (scaled) density perturbation carried by the torsional mode. Hence
\[
\nabla p_i\sim kp_i=k\rho_i c_{Si}^2\frac{\rho_i}{\rho_{0i}}\sim \rho_{0i}c_{Si}^2\epsilon_i\frac{k^2v_i}{\omega},
\]
where $c_{Si}^2=\gamma p_{0i}/\rho_{0i}$ is the square of the adiabatic ion sound speed. The ion-neutral drag force in Equation \ref{eq:A1} has the magnitude
\[
|\rho_{0i}\nu_{in}(v_i-v_n)|\sim \rho_{0i}\nu_{in}\Delta v_i\sim \rho_{0i}\omega v_i,
\]
where we used the scaling $\Delta v_i\sim (\omega/\nu_{in})v_i$, i.e. ions try to move Alfv\'enically while neutrals lag. Therefore, we can define the quantity
\[
{\mathcal R}_i=\frac{\nabla p_i}{|\rho_{0i}\nu_{in}(v_i-v_n)|}\sim \epsilon \frac{c_{Si}^2}{\omega^2}k^2=\epsilon\left(\frac{c_{Si}}{c_A}\right)^2\frac{1}{|Y|^2},
\]
where the dimensionless quantity $Y$ has been defined in the main body of the paper. 

Now let us estimate the value of ${\mathcal R}_i$ under typical partially ionised solar atmospheric conditions. In the figures shown in the main body of the paper, it is clear that in the case of Alv\'en waves, $Y={\mathcal O}(1)$ and the most dissipative regime occurs at $X\sim 1$. Considering $c_{Si}\sim 7-10$ km s$^{-1}$ and $c_A\sim 30-100$ km s$^{-1}$, we obtain $(c_{Si}/c_A)^2\sim 5\times 10^{-3}-10^{-1}$. A plasma will be considered incompressible if ${\mathcal R}_i\ll 1$. Taking a generous $\epsilon=0.1$, we obtain that ${\mathcal R}_i\sim 5\times 10^{-4}-10^{-2}$. This result suggests that the ion pressure gradient is at least two orders of magnitude smaller than the collisional drag (and typically negligible) for torsional Alfv\'en dynamics. This supports omitting $\nabla p_i$ in the charged-fluid equation.

Now, let us turn to the problem of neglecting the neutral gas pressure in the momentum equation for neutrals. In the presence of a pressure gradient force, Equation \ref{eq:2} becomes 
\begin{equation}
\rho_{0n}\left(\frac{\partial}{\partial t}+{\bf v}_{0n}\cdot \nabla\right) \textbf{v}_n=-\rho_{0i}  \nu_{in}(\textbf{v}_n-\textbf{v}_i)-\nabla p_n,
  \label{eq:A10}
\end{equation}
where $p_n$ is the neutral gas pressure perturbation. As in the case of ions, we consider that to leading order the perturbations are incompressible so that $\nabla \cdot {\bf v}_n=0$ and $\rho_n=0$, which leads to $\nabla p_n=0$. We bound the largest possible pressure-gradient contribution by allowing a small residual compressibility, so that $\nabla\cdot {\bf v}_n\neq 0$. In order to cast this, we define the small parameter $\epsilon_n=v_{n\parallel}/v_n$, with $0<\epsilon_n\ll 1$. As in the case of ions, we can write that
\[
\frac{\rho_n}{\rho_{0n}}\sim \frac{v_{n\parallel}}{\omega}=\epsilon\frac{kv_n}{\omega}.
\]
Then, for the neutral pressure gradient, we can write
\[
|\nabla p_n|\sim kp_n\sim k\rho_{0n}c_{Sn}^2\frac{\rho_n}{\rho_{0n}}\sim \rho_{0n}c_{Sn}^2\epsilon_n\frac{k^2v_n}{\omega},
\]
where $c_{Sn}^2=\gamma p_{0n}/\rho_{0n}$ is the square of the neutral sound speed. For the collisional drag force, we can write
\[
|\rho_{0i}\nu_{in}({\bf v}_n-{\bf v}_i)|\sim \rho_{0i}\nu_{in}\Delta v_n\sim \rho_i\omega v_n,
\]
where, again, we used the scaling $\Delta v_n\sim (\omega/\nu_{in})v_n$. Now we can construct the ratio
\[
{\mathcal R}_n=\frac{|\nabla p_n|}{|\rho_{0i}\nu_{in}({\bf v}_n-{\bf v}_i)|}\sim\frac{\rho_{0n}}{\rho_{0i}}\frac{c_{Sn}^2}{\omega^2}k^2\epsilon_n\sim \chi\epsilon_n\left(\frac{c_{Sn}}{c_A}\right)^2\frac{1}{|Y|^2},
\]
where the parameter $\chi$ has been introduced in the main part of the paper. Let us estimate the magnitude of this ratio, taking typical solar chromospheric values. By considering $c_{Sn}=7-10$ km s$^{-1}$, $c_A=70-100$ km s$^{-1}$, we obtain $(c_{Sn}/c_A)^2\sim 5\times 10^{-3}-2\times 10^{-2}$. According to the plots displayed earlier, Alfv\'enic branches have $|Y|\sim {\mathcal O}(1)$. Considering the generous $\epsilon_n=0.1$ and the "worst case" scenario of $\chi=2$, we obtain ${\mathcal R}_n\sim 10^{-3}-4\times 10^{-2}$ that justifies the neglect of the pressure gradient force in comparison with the collisional drag force.

Finally, we need to make a note regarding the applicability of our analysis. At small enough values of the parameter $X$ (low collisionality), a neutral pressure term can reach the same magnitude as the collisional term. In this sense, the obtained relations break down below some critical value of $X$. Let us estimate this threshold. In our study, the pressure gradient is omitted (i.e., no $-\nabla p_n$ term). This assumption is only valid if the collisional drag dominates over the pressure gradient in the neutral dynamics. Let us estimate the threshold when this statement holds. The collisional drag force scales as $F_c\sim \rho_n\nu_{in}v$, while the pressure force can be written as $F_p\sim p_n/L\sim\rho_nc_{Sn}^2k^2v/\omega$, where $L$ is a characteristic length scale related to waves and $v$ is the characteristic speed of the neutral species. For simplicity, we take $L$ to be proportional to the wavelength of waves, therefore $L\sim 1/k$. As a result, the ratio of the two forces can be written as
\[
\frac{F_p}{F_c}\sim \frac{c_{Sn}^2k^2}{\omega\nu_{in}}=\frac{c_{Sn}^2}{c_A^2}\frac{1}{XY}
\]
Pressure forces can be neglected when $F_p/F_c\ll 1$, so when $X\gg c_{Sn}^2/c_A^2Y$. Choosing $Y\sim {\mathcal O(1)}$, our analysis is valid as long as the dimensionless collisional parameter satisfies the condition $X\gg c_{Sn}^2/c_A^2$. Taking, for example, $c_{Sn}=10$ km s$^{-1}$ and $c_A=100$ km s$^{-1}$ as characteristic speeds for the partially ionised solar chromosphere, we obtain that pressure forces become important when $X\leq 10^{-2}$.

%%% BIBLIOGRAPHY %%%%%%%%%%%%%%%%%%%%%%%%%%%%%%%%%%%%%%%%%%%%%%%%%%%%%%%%%%%

     % format of references provided by the journal (.bst)
\bibliographystyle{spr-mp-sola}
     % name your Bibtex file containing your references (.bib)
\bibliography{sola_bibliography_example}  

     % Checking: look if the file containing the ``\bibitem'' exits
     %           so check if the .bbl file exist (bibTeX compilation)
%\IfFileExists{\jobname.bbl}{} {\typeout{}
\typeout{****************************************************}
\typeout{****************************************************}
\typeout{** Please run "bibtex \jobname" to obtain} \typeout{**
the bibliography and then re-run LaTeX} \typeout{** twice to fix
the references !}
\typeout{****************************************************}
\typeout{****************************************************}
%\typeout{}}

\end{document}